  \def\draftversion{false}
\documentclass[prb,preprint,showpacs,amsmath,amssymb,superscriptaddress]{revtex4-2}
\usepackage{footnote} % footnotes
\usepackage{graphicx}% Include figure files
\usepackage{subcaption}
\usepackage{lipsum} % for dummy text
\usepackage{wrapfig}% Have to use a package to do text wrapping
\usepackage{ifthen}% To use conditional command ifthenelse
\usepackage{color} % So that I can color the comment marks
\usepackage{bm} % bold math
\usepackage{multirow}% For tables
\usepackage{hyperref} % adds links
\usepackage{bbold} % allow for unit matrix symbol
\hypersetup{backref=true,
 pdfnewwindow=true, colorlinks=true,
 linkcolor=blue, anchorcolor=blue,
 citecolor=blue, filecolor=blue,
 menucolor=blue, urlcolor=blue}
\ifthenelse{\equal{\draftversion}{true}}{
  \marginparwidth 2.7in
  \marginparsep 0.5in
  \newcounter{comm} % counter for commentaries
  \def\commnext{\stepcounter{comm}}
  \def\commtext{{\bf\color{blue}[\arabic{comm}]}}
  \def\commmar{{\bf\color{blue}[\arabic{comm}]}}
  \def\jlm#1{\commnext\marginpar{\small JL\commmar: #1}\commtext}
  \def\jdm#1{\commnext\marginpar{\small JRD\commmar: #1}\commtext}
  \def\bbm#1{\commnext\marginpar{\small BAB\commmar: #1}\commtext}
}{
  \def\jlm#1{}
  \def\jdm#1{}
  \def\bbm#1{}
}
\def\nn{\nonumber\\}
\def\vr{\mathbf{r}}

\def\vq{\mathbf{q}}

\def\xc{\textrm{xc}}
\def\beq{\begin{equation}}
\def\eeq{\end{equation}}

\def\nn{\nonumber\\}

\def\vr{\mathbf{r}}

\def\vk{\mathbf{k}}
\def\vG{\mathbf{G}}
\def\vq{\mathbf{q}}

\def\bsp{\begin{split}}
\def\esp{\end{split}}
\def\nn{\nonumber\\}

\def\xc{\textrm{xc}}

\def\Bi2Se3{\textrm{Bi}_2\textrm{Se}_3}

\begin{document}

\title{ The electronic structure of $\beta$-HgS via $GW$ calculations}

\author{Bradford A. Barker}
\affiliation{Department of Physics, University of California, Merced, CA 95343, USA}
\affiliation{Department of Physics, University of California, Berkeley, CA 94720, USA}
\affiliation{Materials Sciences Division, Lawrence Berkeley National Laboratory, Berkeley, CA 94720, USA}
\author{Steven G. Louie}
\affiliation{Department of Physics, University of California, Berkeley, CA 94720, USA}
\affiliation{Materials Sciences Division, Lawrence Berkeley National Laboratory, Berkeley, CA 94720, USA}

\date{\today}


\begin{abstract}
The electronic structure of the zincblende $\beta$-HgS is not well understood. Previous first-principles calculations using fully-relativistic density functional theory and many-body perturbation theory in the fully-relativistic $GW$ approach have predicted an inverted, topologically non-trivial ordering of these states, with the $s$-like $\Gamma_6$ state occupied.
However, other calculations using the $GW$ approach in which spin-orbit coupling is added perturbatively (``$GW$+SOC'')
predict the $p$-$d$ hybridized $\Gamma_7$ and $\Gamma_8$ states to be occupied and the $\Gamma_6$ state to be unoccupied, suggesting that $\beta$-HgS is a topologically trivial small band gap semiconductor.
In the present work, a plane-wave pseudopotential fully-relativistic $GW$ calculation finds a band ordering in agreement with the previous $GW$+SOC calculations. The calculated band gap is 0.10~{eV} and the electron effective mass is 0.07~$m_e$, in good agreement with experiment.
\end{abstract}

\maketitle

\section{Introduction}
\label{sec:intro}

Metacinnabar, or $\beta$-HgS, has a zincblende structure and large spin-orbit coupling.
The related mercury chalcogenide zincblende solids HgSe and HgTe have a semimetallic, $\alpha$-Sn-like bandstructure, with parabolic valence and conduction bands degenerate at the $\Gamma$-point.
For $\alpha$-Sn, HgSe, and HgTe, the bands near the Fermi level at the $\Gamma$-point have an inverted
bandstructure: The order of states, in increasing energy, for a conventional zincblende system is the $p_{1/2}$-like $\Gamma_{7}$, $p_{3/2}$-like $\Gamma_{8}$, and $s$-like $\Gamma_{6}$,
while HgSe and HgTe place $\Gamma_6$ lowest in energy, then $\Gamma_7$, then $\Gamma_8$.
Furthermore, the four-fold degenerate $\Gamma_8$ splits into a pair of parabolic valence and conduction bands, degenerate at the $\Gamma$-point. This behavior in the bandstructure for these materials is in agreement in the literature, regardless of the details by which spin-orbit coupling is incorporated.\cite{rohlfing_hgse,van_schilfgaarde,sakuma}

The bandstructure for $\beta$-HgS, however, has seen considerable disagreement between different theoretical calculations and experiment.
The $\alpha$-Sn-like inverted bandstructure was proposed to be consistent with experimental measurements \cite{zallen}.
Reflectivity data indicates a plasma edge at 0.10~eV\cite{zallen}, suggesting a metallic or semimetallic nature.
Absorption data indicated an onset of interband transitions at 0.25~eV,
%(after free-carrier-like absorption),
interpreted as the onset of transitions to a partially occupied, zero-gap parabolic conduction band\cite{zallen}.
In the absence of ARPES spectra, however, there is not a definitive experimental description of the quasiparticle bandstructure.
%and the true ordering of the states.

Density functional theory calculations\cite{hohenberg_kohn,kohn_sham}, however, indicate a small indirect band gap and
states ordered $\Gamma_{6}$, $\Gamma_{8}$, $\Gamma_{7}$\cite{delin}.
Compared to HgSe and HgTe, there is a further inversion between $\Gamma_{8}$ and $\Gamma_7$,
with $\Gamma_8$ now being fully occupied even away from the $\Gamma$-point. 
%The semimetallic state
%is only recreated in DFT with the neglect of spin-orbit coupling (SOC).
%As the spin-orbit strength within the Hg 5$d$ subshell is 1.86~{eV}\cite{hg_atomic}, spin-orbit coupling is too large to neglect.
Based on the fully-relativistic DFT bandstructure, $\beta$-HgS has been predicted to be a
nontrivial $Z_2$ insulator much like strained HgTe, with highly anisotropic topologically protected Dirac surface states along the [001]
direction\cite{ti,tss}. Topological properties based on the ordering of bands near the Fermi energy need to be confirmed
from calculations that are more accurate than DFT\cite{zunger}, which is well-known to underestimate band gaps.
Many-body perturbation theory in the $GW$ approximation provides physically accurate excited-state properties such as the electronic bandstructure.
$GW$ calculations in which the spin-orbit coupling Hamiltonian is applied as a perturbation to the quasiparticle energies (``$GW$+SOC'')
predict a band ordering
of $\Gamma_{8}$, $\Gamma_{7}$, $\Gamma_{6}$\cite{fleszar,van_schilfgaarde},
which is similar to that of CdTe, but with the $p$-$d$ hybridized orbitals
$\Gamma_{8}$ and $\Gamma_{7}$ inverted, due to the strength of SOC in the Hg 5$d$ states. This ordering of states yields a topologically trivial band gap, as the bandstructure can be adiabatically
deformed to that of the topologically trivial CdTe\cite{bernevig}
without closing the bulk band gap by, e.g., tuning the atomic spin-orbit parameters of Hg and S.
%A fully-relativistic DFT calculation with an additional semiempirical local potential parameterized from experimental energy levels at the high-symmetry
%Brillouin Zone points for HgTe also yielded an ordering of $\Gamma_{8v}$, $\Gamma_{7v}$, $\Gamma_{6c}$, with a (non-inverted)
%band gap of 0.30 eV for $\beta$-HgS\cite{wei}. This local potential can be regarded as an empirical approximation to the self-energy corrections
%to a DFT calculation.

Previous $GW$ calculations in which spin-orbit coupling is incorporated non-perturbatively through the use of fully-relativistic pseudopotentials for the Kohn-Sham equations and spinor wavefunctions in the construction of the self-energy (``FR-$GW$'') yield
the same band ordering as in DFT\cite{sakuma,west}.
%The computed quasiparticle energy difference between the $\Gamma_6$ and $\Gamma_8$ states is small when using an LDA exchange-correlation functional as the mean-field starting point\cite{sakuma}, and a slightly larger energy difference when calculated using the GGA exchange-correlation functional\cite{west}.

The disagreement in the various computed quasiparticle bandstructure topologies is curious, since the fully-relativistic DFT (``FR-DFT'')
bandstructures are all in agreement despite the different choices of basis sets.
In this work, the FR-$GW$ quasiparticle bandstructure is calculated using a plane-wave basis and approximating the electron-ion interaction within
the Local Density Approximation (LDA)\cite{perdew_zunger} when constructing fully-relativistic pseudopotentials for the Kohn-Sham wavefunction calculations. These wavefunctions are used as the basis for the quasiparticle bandstructure calculations.
In Section \ref{sec:methods} the method of ``one-shot'' FR-$GW$ in a plane-wave basis is reviewed, and 
in Section \ref{sec:electronic_structure} the electronic structure is discussed.

%the states at the $\Gamma$-point are ordered, in increasing energy, $\Gamma_8$, $\Gamma_7$, then $\Gamma_6$.

%%========================================================================
\section{Methods}
\label{sec:methods}
%%========================================================================

The Hedin equations with the inclusion of spin-orbit coupling, within the $GW$ approximation, are\cite{biermann}

\begin{align}
& W(\vr_1,\vr_2;\omega) = v(\vr_1,\vr_2)\, + \, \int \textrm{d}\vr_3 \, \textrm{d}\vr_4 \, \textrm{d}\omega ' \, v(\vr_1,\vr_3) \, P(\vr_3,\vr_4;\omega ' )\, W(\vr_4,\vr_2;\omega - \omega '), \nn
& P(\vr_1,\vr_2) = -\frac{i}{2\pi} \sum_{s_1,\, s_2 }\int \textrm{d}\omega ' \, G(\vr_1,s_1,\vr_2,s_2;\omega + \omega ') \, G(\vr_2,s_2,\vr_1,s_1;\omega '), \nn
& \Sigma(\vr_1,s_1,\vr_2,s_2;\omega) = \frac{i}{2\pi} \int \textrm{d}\omega ' e^{-i0^{+}\omega '} \, G(\vr_1,s_1,\vr_2,s_2; \omega-\omega ') \, W(\vr_1,\vr_2; \omega '), \nn
& \Gamma(\vr_1,s_1,\vr_2,s_2,\vr_3) = \delta(\vr_1-\vr_2)\delta(\vr_2-\vr_3)\delta_{s_1,s_2}\, 
\end{align}
with the one-particle Green's function constructed using Kohn-Sham energies, $\epsilon_{n\vk}$, and orbitals\cite{hybertsen,sakuma}, $\phi_{n\vk\alpha}(\vr)$,
\begin{equation}
G\left(\vr_1,s_1,\vr_2,s_2; \omega\right) \approx \sum_{n\vk} \frac{\phi_{n\vk}(\vr_1,s_1)\phi^{*}_{n\vk}(\vr_2,s_2)}{\omega - \epsilon_{n\vk}-i\delta_{n\vk}},
\end{equation}
where $\delta_{n\vk} = 0^{+}$ for $\epsilon_{n\vk} < \mu$ and $\delta_{n\vk} = 0^{-}$ for $\epsilon_{n\vk} > \mu$.

The Kohn-Sham energies and orbitals are calculated using \texttt{Quantum ESPRESSO}\cite{espresso}, with fully-relativistic pseudopotentials for Hg and S generated from the Optimized Norm-Conserving Vanderbilt Pseudopotential method\cite{oncvpsp} with parameters adapted from the Pseudo-Dojo pseudopotential database\cite{pseudo_dojo,pseudo_dojo_reproducibility}.
The Hg pseudopotential includes the
$5s^{2}5p^{6}5d^{10}$
semicore states as valence for accurate calculation of the bare exchange matrix elements
\cite{semicore}.
A kinetic energy cutoff of 200 Ry and an 8$\times$8$\times$8 Monkhorst-Pack grid are used for calculating the charge density and the relaxed structural geometry. The relaxed lattice parameter is calculated to be identical to the experimental value of 5.85~{\AA} \cite{madelung}.

The calculation for the polarizability is identical to the case in which spin-orbit coupling is neglected
apart from the calculation of the plane-wave matrix elements (and any differences in the Kohn-Sham eigenvalues), where the spin degree of freedom appears as a trace:

\begin{equation}
M_{nn'}(\vk,\vq,\vG) = \sum_{s}\langle n,\, \vk+\vq,\, s\,|e^{i(\vq+\vG)\cdot \vr} |n',\, \vk,\, s \rangle,
\end{equation}
\begin{align}
& P_{\vG\vG'}(\vq,\omega) = \sum\limits_{n}^{\textrm{occ}}\sum\limits_{n'}^{\textrm{unocc}}\sum_{\vk}
M^{*}_{nn'}(\vk,\vq,\vG)M_{nn'}(\vk,\vq,\vG') \nonumber \frac{1}{2}\left[\frac{1}{\epsilon_{n\vk+\vq} - \epsilon_{n'\vk}-\omega + i\delta}
+ \frac{1}{\epsilon_{n\vk+\vq} - \epsilon_{n'\vk}+\omega + i\delta}\right].
\end{align}

The quasiparticle eigenvalues are the poles of Dyson's equation, which in the usual approximation where the Kohn-Sham orbitals are taken to be the quasiparticle orbitals gives
\begin{equation}
E_{n\vk} = \epsilon_{n\vk} + \sum_{s_1,s_2}\langle n,\, \vk,\, s_1 |\left( \Sigma\left(s_1,s_2,E_{n\vk}\right) - V^{\textrm{xc}}\delta_{s_1,s_2}\right) | n,\, \vk,\, s_2 \rangle.
\end{equation}
(The spatial dependence of $\Sigma$ has been suppressed for brevity.) The calculation of the matrix elements of the self-energy is identical to that of the spinless case\cite{BerkeleyGW}
apart from the trace over spin in the calculation of the plane-wave matrix elements $M_{nn'}$.
In the usual screened-exchange/Coulomb-hole partitioning of the self-energy\cite{hybertsen},
the self-energy matrix elements\cite{BerkeleyGW, catalin_thesis} to evaluate are
\begin{align}
\sum_{s_1,s_2} \langle n,\,\vk,\,s_1 | \Sigma^{\textrm{COH}}(s_1,s_2;\omega) | m,\,\vk,\, s_2 \rangle = \frac{i}{2\pi}&\sum\limits_{n'} \sum\limits_{\vq\vG\vG'}M^{*}_{n'n}(\vk,-\vq,-\vG)M_{n'm}(\vk,-\vq,-\vG')\nn
&\times\int \textrm{d}\omega' \, \frac{\Im\,\epsilon^{-1}_{\vG\vG'}(\vq;\omega')}{\omega - E_{n'\vk-\vq}-\omega' + i\delta} \, v(\vq + \vG'),\nn
\end{align}
\begin{align}
& \sum_{s_1,s_2}\langle n,\,\vk,\,s_1 | \Sigma^{\textrm{SEX}}(s_1,s_2;\omega) | m,\,\vk,\,s_2 \rangle =\nn
& \;\;\;\; -\sum\limits^{\textrm{occ}}_{n'} \sum\limits_{\vq,\vG,\vG'}M^{*}_{n'n}(\vk,-\vq,-\vG)M_{n'm}(\vk,-\vq,-\vG')\epsilon^{-1}_{\vG\vG'}(\vq;\omega) v(\vq + \vG').
\end{align}
In this work, the polarizability is evaluated by explicitly calculating
using the contour deformation method\cite{contour_deformation} with the static subspace approximation\cite{low_rank,govoni_galli,mauro} with the \texttt{BerkeleyGW}\cite{BerkeleyGW} excited-state code modified for use with spinor wavefunctions.
The dielectric matrix cutoff is 35 Ry, and 2000 unoccupied states
are used for both the polarizability and Coulomb-hole sums. The error in the band gap due to the use of 2000 empty states in the Coulomb-hole sum
is estimated to be 7~{meV}. An 8$\times$8$\times$8 q-point grid for the dielectric function is used, as this grid has been shown to be sufficient
for accurate calculations of the band gap for the similarly sized diamond-structure material, Ge\cite{malone}.
%These calculations are expedited by the use of the static subspace method\cite{low_rank,govoni_galli,mauro},
%in which we find the eigenvectors of $\epsilon_{\vG\vG'}(\vq, \, \omega = 0)$ for each $\vq$-point and use a fraction of that basis in the calculation of the self-energy matrix elements.
For the static subspace approximation, the lowest 10 percent of the static dielectric matrix eigenvectors ($N_{\textrm{basis}}=117$) gives a converged fundamental band gap within 6~{meV}.
15 imaginary frequencies for the contour deformation method give a band gap converged within 3~{meV}, compared to the use of 25 frequencies.
The $\vq\rightarrow 0$ limit is treated by the dual grid technique, appropriate
for a semiconductor even with a small gap\cite{BerkeleyGW}. The quasiparticle energies calculated by the Hybertsen-Louie Generalized Plasmon Pole model\cite{hybertsen} (``GPP'') and the contour deformation method include the contributions from the static-remainder method\cite{static_remainder} to estimate the correction from the missing bands from the finite Coulomb-hole sum.
%%========================================================================
\section{Electronic Structure}
\label{sec:electronic_structure}
%%========================================================================

The conventional band gap for a zincblende semiconductor is defined to be
$E_0 = E(\Gamma_6)-E(\Gamma_8)$, which is positive for usual zincblende materials but negative
for systems with band inversion.
%For the usual zincblende materials (e.g. GaAs or CdTe), this is equal to the true band gap.
Likewise, the spin-orbit splitting $\Delta^{\textrm{SOC}} = E(\Gamma_8)-E(\Gamma_7)$ is also
defined to be positive for the usual zincblende materials. However, the Hg 5$d$ orbitals contribute to these states significantly,
as $p$-$d$ orbital hybridization is allowed for tetrahedral symmetries\cite{weiprb}.
%The spin-orbit coupling of these states, then, is dominated by the contribution of the Hg 5$d$ atomic states.
Since the Hg 5$d_{5/2}$ state contributes to the $\Gamma_7$
and the Hg 5$d_{3/2}$ to $\Gamma_8$, $\Delta^{\textrm{SOC}}$ will be negative, as the spin-orbit splitting in these states (1.86~{eV}\cite{hg_atomic})
is much larger than that of the S 3$p_{3/2}$ and 3$p_{1/2}$ states ($< 0.10$~{eV}\cite{s_atomic}).

\begin{figure}[p]
  \centering
  \includegraphics[width=0.5\textwidth]{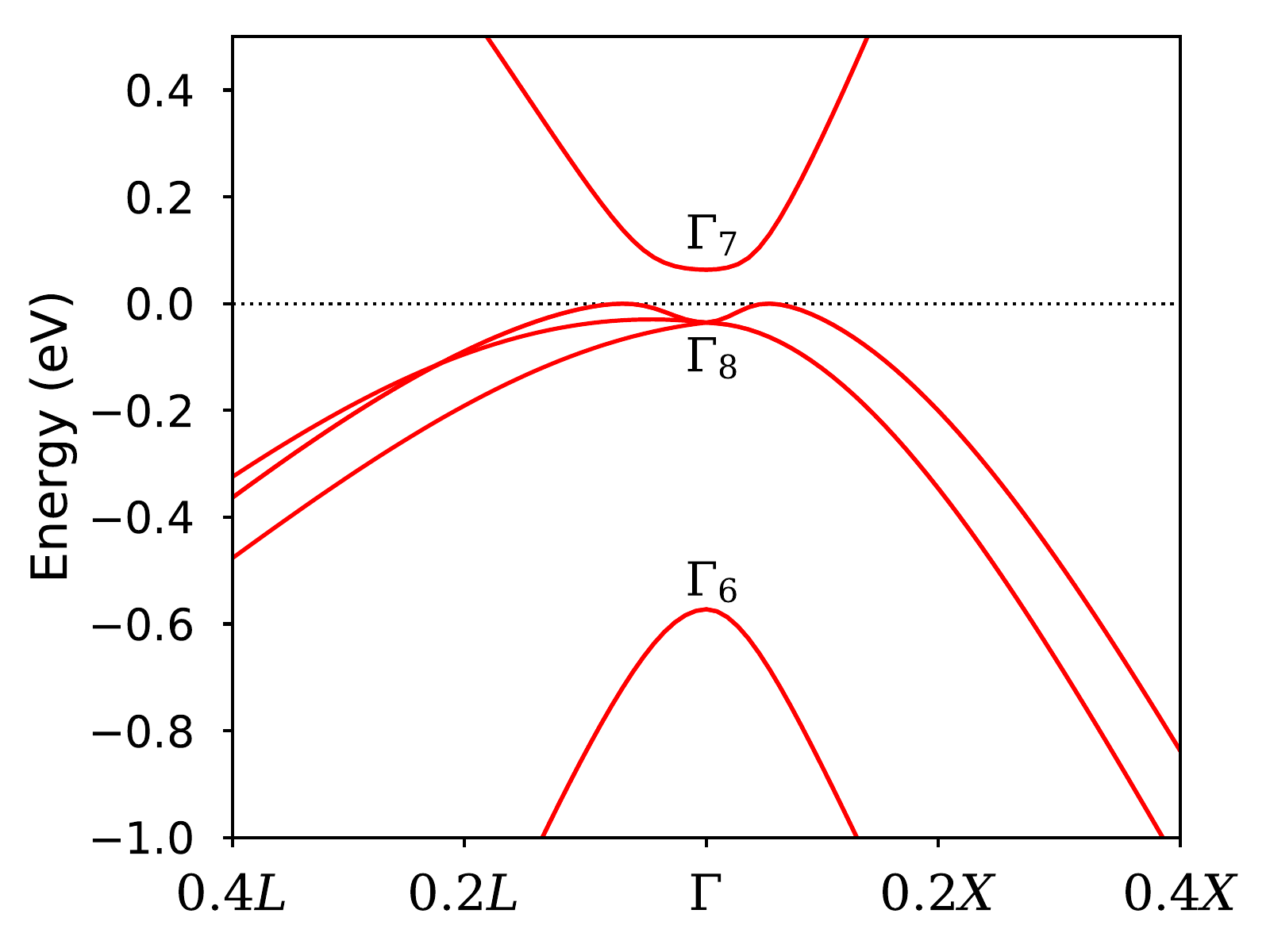}
  \caption{The fully-relativistic DFT bandstructure of $\beta$-HgS. The states at the $\Gamma$-point, in increasing energy, are $\Gamma_6$, $\Gamma_8$, $\Gamma_7$.}
  \label{fig:rel_dft}
\end{figure}

The fully-relativistic DFT bandstructure is shown in Fig. \ref{fig:rel_dft}. At $\Gamma$, spin-orbit coupling breaks the degeneracy at the Fermi level and opens a small
spin-orbit gap, with the $\Gamma_7$ states unoccupied and the four-fold $\Gamma_8$ states occupied. The small indirect gap is 0.10~{eV}.

The quasiparticle energies within FR-$GW$ approach are calculated for the $\Gamma$, 1/8 $L$, 1/4 $L$, 1/2 $L$, 3/4 $L$, $L$, 1/8 $X$, 1/4 $X$, 1/2 $X$, 3/4 $X$, and $X$-points.
%treating the frequency-dependence of the self-energy operator by the contour deformation method and the static subspace approximation.
The bandstructure is then plotted with cubic splines, as the usual approach in which the quasiparticle bandstructure is linearly interpolated from the DFT bandstructure\cite{BerkeleyGW} requires the DFT bandstructure to be qualitatively similar to the quasiparticle bandstructure\cite{BerkeleyGW}.

The states near the Fermi energy are ordered $\Gamma_{8}$, $\Gamma_{7}$, then $\Gamma_{6}$, in agreement with other $GW$ calculations
in which spin-orbit coupling is added perturbatively\cite{fleszar,van_schilfgaarde}.
This is in contrast to the other reported FR-$GW$ calculations in the literature\cite{sakuma,west}, where the
states are ordered, as in DFT, $\Gamma_{6}$, $\Gamma_{8}$, then $\Gamma_{7}$. 
%We note that the authors in Ref. \onlinecite{sakuma} constructed their polarizability by using the states and eigenvalues from a DFT calculation ignoring spin-orbit coupling. The bandstructure in this case has a vanishing energy gap at $\Gamma$, in contrast to the small gap when including SOC. The vanishing gap will yield a divergence in the dielectric function for $\vq \rightarrow 0$, giving an overestimation of screening and a smaller correction to the band gap than would be otherwise given.
%The calculation of the band gap and electron effective mass were found to agree well with the literature.
The experimental band gap was estimated to be -0.11~{eV} from a Shubnikov-de Haas measurement of transition metal-doped samples
of $\beta$-HgS, with the carrier concentration extrapolated to zero\cite{shubnikov}. The model used to interpret the experiment, however,
depends only on the absolute value of the band gap\footnote{This was first noted in Ref. \onlinecite{van_schilfgaarde}}. With the view that the experimental band gap may be positive,
the calculated value of 0.10~{eV} is in good agreement.
Similarly, the calculated electron effective mass to be 0.07~$m_e$. This is in the range of values cited
for the transition metal-doped $\beta$-HgS samples, 0.04~$m_e$ to 0.07~$m_e$\cite{shubnikov}.
The electron effective mass estimated from reflectivity measurements also gives the value 0.07~$m_e$\cite{zallen}.
The gap at $\Gamma$ has also been calculated
with the frequency dependence of the self-energy treated by the COHSEX and Hybertsen-Louie Generalized Plasmon Pole model\cite{hybertsen,complex_gpp} approximations, which give different values for the band gap but give the same order of the states. The band gaps are summarized in Table \ref{tab:table1}. 
%, in addition to explicit calculation of the frequency dependence via the contour deformation method.
%The (one-shot) Hartree-Fock gap is determined from the COHSEX matrix elements with the
%static correlation term in the COHSEX self-energy subtracted out.
%The topologically trivial ordering of bands was reproduced
%within COHSEX and GPP FR-$GW$.
%The ordering of states was indeed the same in all cases.

%Since even the Hartree-Fock energies place the bands in the topologically trivial ordering,
%it appears that the calculation of the bare exchange energy is necessary to determine the %qualitative features of the quasiparticle bandstructure.
%With the use of all of the subshells in the fifth atomic shell for the Hg psuedopotential, we help %ensure that the exchange energy
%is accurately calculated\cite{semicore}.

To confirm the ordering of the states, off-diagonal COHSEX calculation was performed to estimate the degree by which the quasiparticle wavefunctions differ from the Kohn-Sham orbitals.
% accurately describe the quasiparticle wavefunctions, apart from a mere reordering of states.
%Care must be taken for small gap semiconductors with large spin-orbit coupling that the self-energy is calculated with the appropriate basis set, especially when the ordering of the quasiparticle energies changes relative to LDA.
%
The COHSEX approximation is a static approximation to the self-energy operator, and its frequency-independence allows for a completion relation
to remove the sum over empty states in the Coulom-hole term in the self-energy:
\begin{widetext}
\begin{align}
& \langle n,\, \vk,\, s_1 | \Sigma^{\textrm{COH}}  (s_1,s_2;\omega = 0) | m\vk\beta \rangle = \nn
& \;\;\;\; \frac{1}{2} \sum\limits_{\vq\vG\vG'} \langle n,\,\vk,\,s_1 | e^{i(\vG' - \vG)\cdot \vr}\delta_{s_1,s_2} | m,\,\vk,\, s_2 \rangle\left[ \epsilon^{-1}_{\vG\vG'}(\vq;\omega = 0) - \delta_{\vG\vG'}\right] v(\vq + \vG').
\end{align}
\end{widetext}
Thus COHSEX allows for rapid calculation of matrix elements of an approximate form of the self-energy operator.
%With the matrix elements $\Delta \Sigma_{nm\vk} = \langle n\vk \alpha | \Sigma^{COHSEX}_{\alpha\beta} - V^{xc}_{\alpha\beta} | m\vk\beta \rangle $,
%with $n \neq m$, we can estimate the $m$'th $LDA$ wavefunction's contribution to the first-order correction to the COHSEX wavefunctions $\psi^{(1)}_{n\vk}(\vr)$ from

\begin{figure}[p]
  \centering
  \includegraphics[width=0.5\textwidth]{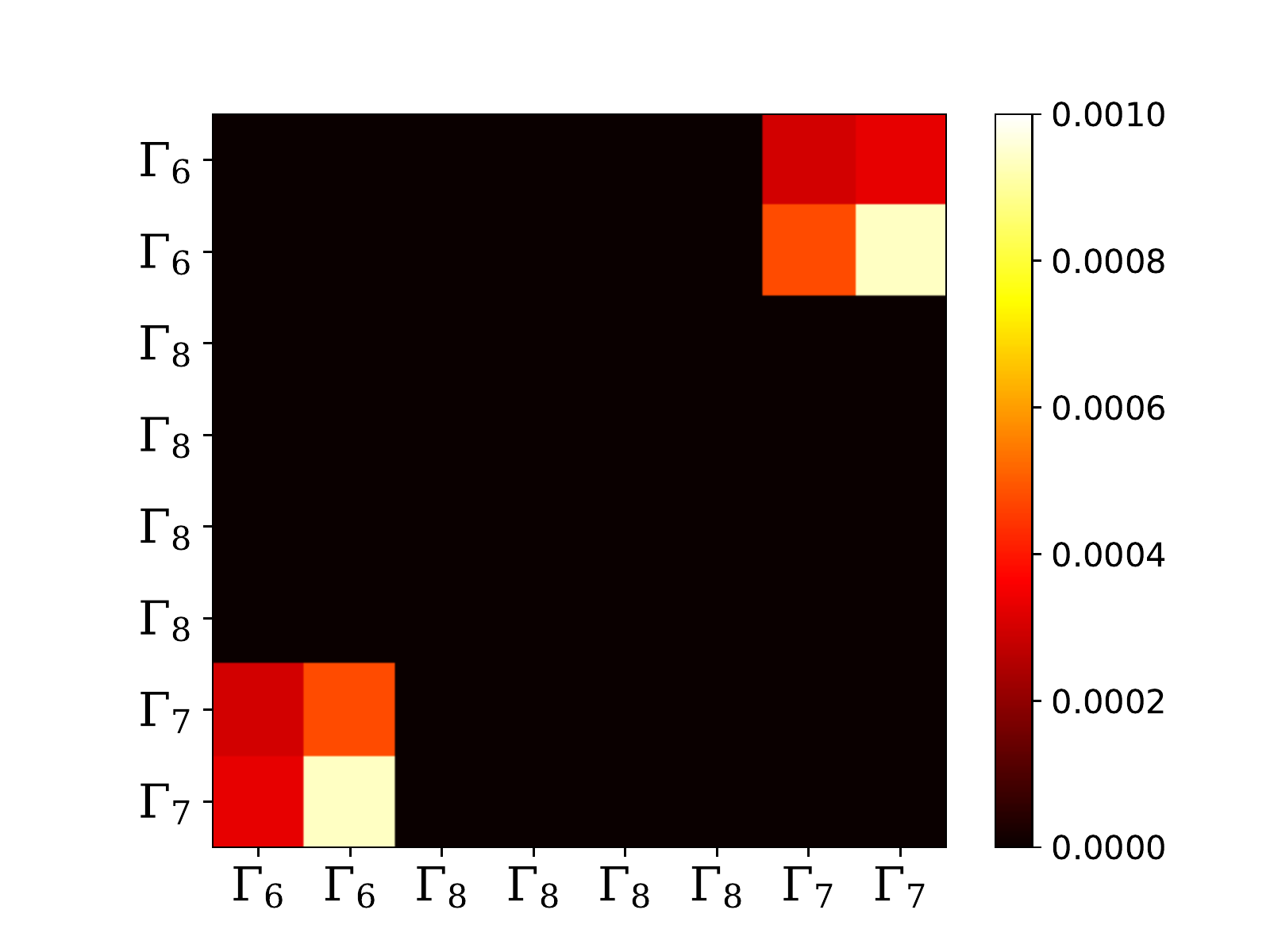}
  \caption{
  %$U^{(1)}_{nm\vk}$
  The first-order correction to each COHSEX wavefunction (rows), within the Kohn-Sham orbital basis (columns).
%           The $n$'th COHSEX state (columns) is the $n$'th LDA state, plus the sum of each contribution of the remaining LDA states (rows),
%           the magnitude of which is indicated by the heat-map. 
           The maximum off-diagonal contribution is less than 0.1 percent.}
  \label{fig:psi1}
\end{figure}

The contribution of an Kohn-Sham orbital $\phi^{\textrm{KS}}_{n\vk}(\vr)$ to the first-order correction to the COHSEX wavefunction $\psi^{(1)}_{n\vk}(\vr)$ is calculated from
\begin{align}
& \psi^{(1)}_{n\vk}(\vr) = \sum\limits_{m\neq n} U^{(1)}_{nm\vk} \phi^{\textrm{KS}}_{m\vk}(\vr),\\
& U^{(1)}_{nm\vk} = \frac{\langle n\vk \alpha | \Sigma^{\textrm{COHSEX}}(s_1,s_2) - V^{\xc}\delta_{s_1,s_2} | m,\,\vk,\,s_2 d\rangle}{\epsilon_{n\vk} - \epsilon_{m\vk}}.
\end{align}
%
%with ``$\Sigma$'' understood to be the COHSEX form of the electron-electron self-energy.
%Nonzero values of $|U^{(1)}_{nm\vk}|$ indicate a contribution from the $m$'th LDA band to the $n$'th static-COHSEX band and we must fully diagonalize
%the static-COHSEX Hamiltonian to arrive at the appropriate wavefunctions.
%In a rather extreme case, the values of $U$ may be distributed in such a manner that the quasiparticle states are re-ordered LDA states. We look, then, for any possible further reordering of the $\Gamma$-point states after the inclusion of off-diagonals.
Fig. \ref{fig:psi1} displays the values of the coefficients $|U^{(1)}_{nm\vk}|$. 
%with the rows $n$ and columns $m$.
Only the $\Gamma_6$ and $\Gamma_7$ states
show a contribution to the others' COHSEX wavefunctions to first order.
However, the largest contribution is less than 0.1 percent of the zero-order contribution, so the Kohn-Sham states are indeed good approximations to the quasiparticle states.

%The $s$-state in LDA is attracted to the Hg potential more strongly due to incomplete screening of the Hg potential from the Hg $d$ electrons.
%With GW corrections, the $d$-like states are more strongly bound and can more effectively screen the Hg ionic potential, and the S $s$-states
%can now be unoccupied. The large spin-orbit coupling of the Hg $d$ states in the $pd$-hybridized $\Gamma_7$ and $\Gamma_8$ allow for the
%$\Gamma_7$ states to be placed higher in energy, as they correspond the $d_{5/2}$ states, while $\Gamma_8$ corresponds to the lower-energy $d_{3/2}$ states.

\begin{figure}[p]
  \centering
  \includegraphics[width=0.5\textwidth]{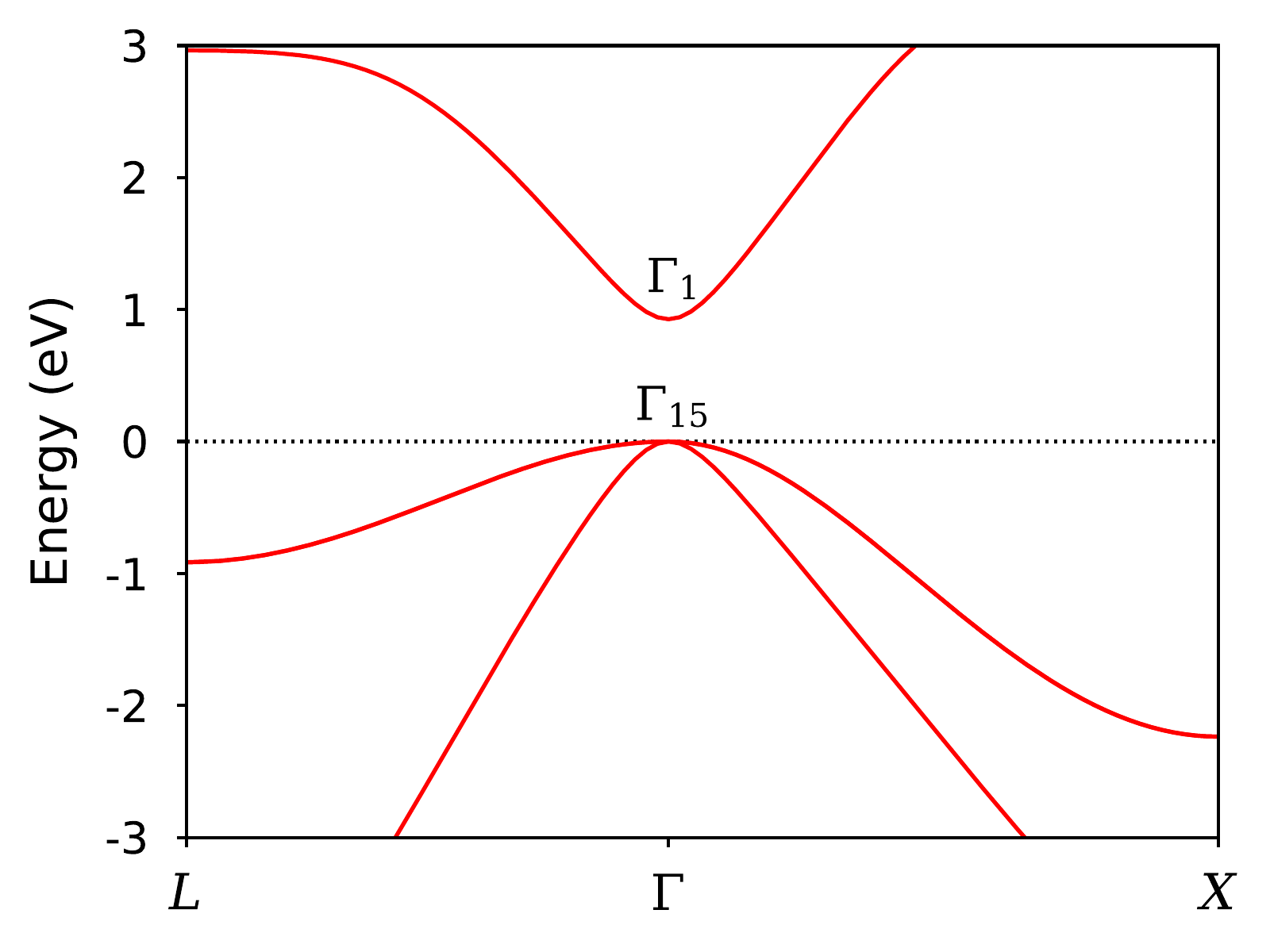}
  \caption{The non-relativistic DFT bandstructure of $\beta$-HgS. The states at the $\Gamma$ point are ordered $\Gamma_{15}$, $\Gamma_1$, as in a conventional zincblende structure.}
  \label{fig:nonrel_dft}
\end{figure}

\begin{figure}[p]
  \centering
  \includegraphics[width=0.5\textwidth]{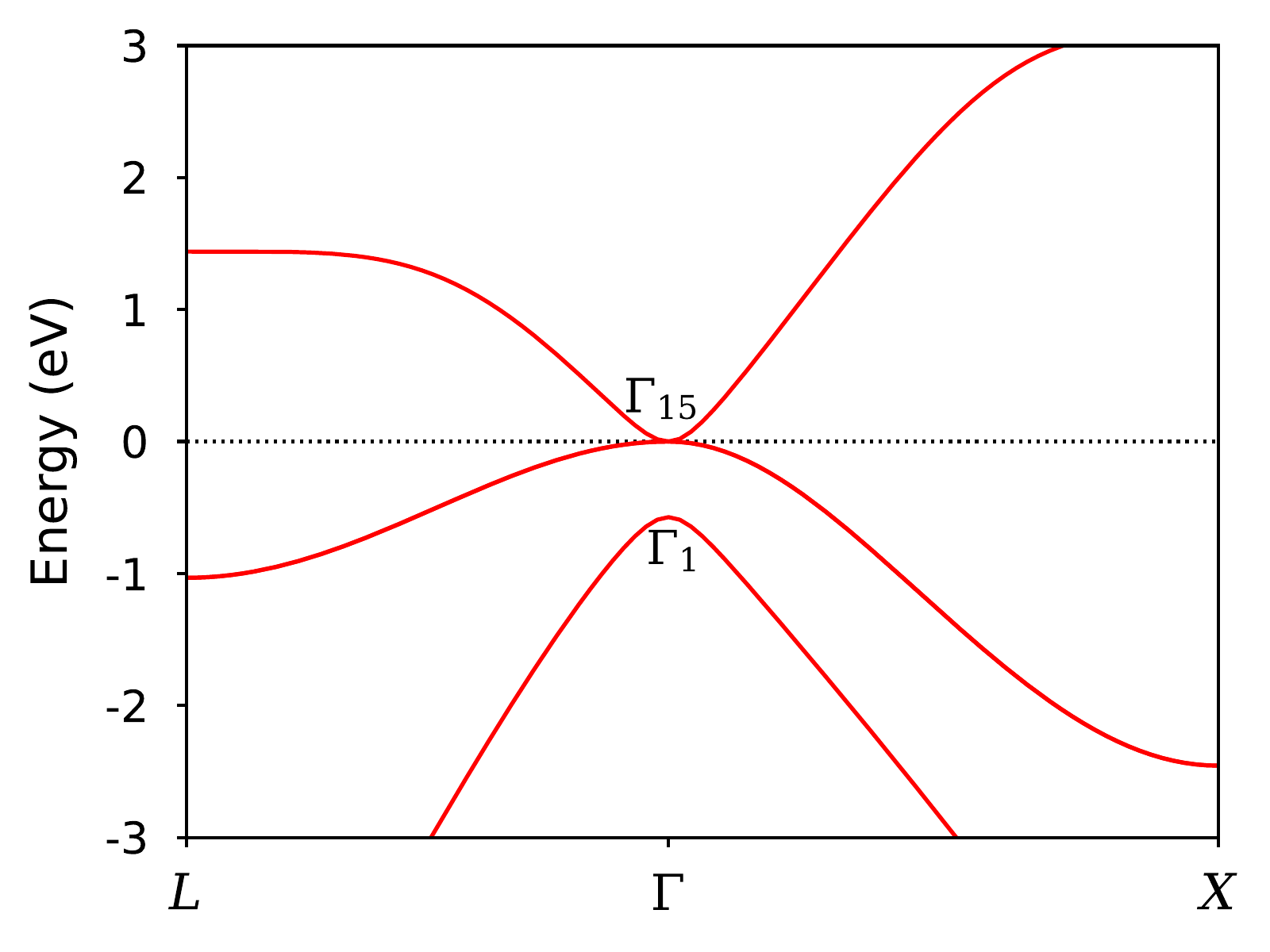}
  \caption{The scalar-relativistic DFT bandstructure of $\beta$-HgS. The degenerate states at $E_\textrm{F}=0$ belong to the $\Gamma_{15}$ representation, and the lower occupied state belongs to $\Gamma_1$.}
  \label{fig:scalar_dft}
\end{figure}

\begin{figure*}[p]
  \centering
    \begin{subfigure}[t]{0.49\textwidth}
    \centering
    \caption{}   
    \includegraphics[width=\textwidth]{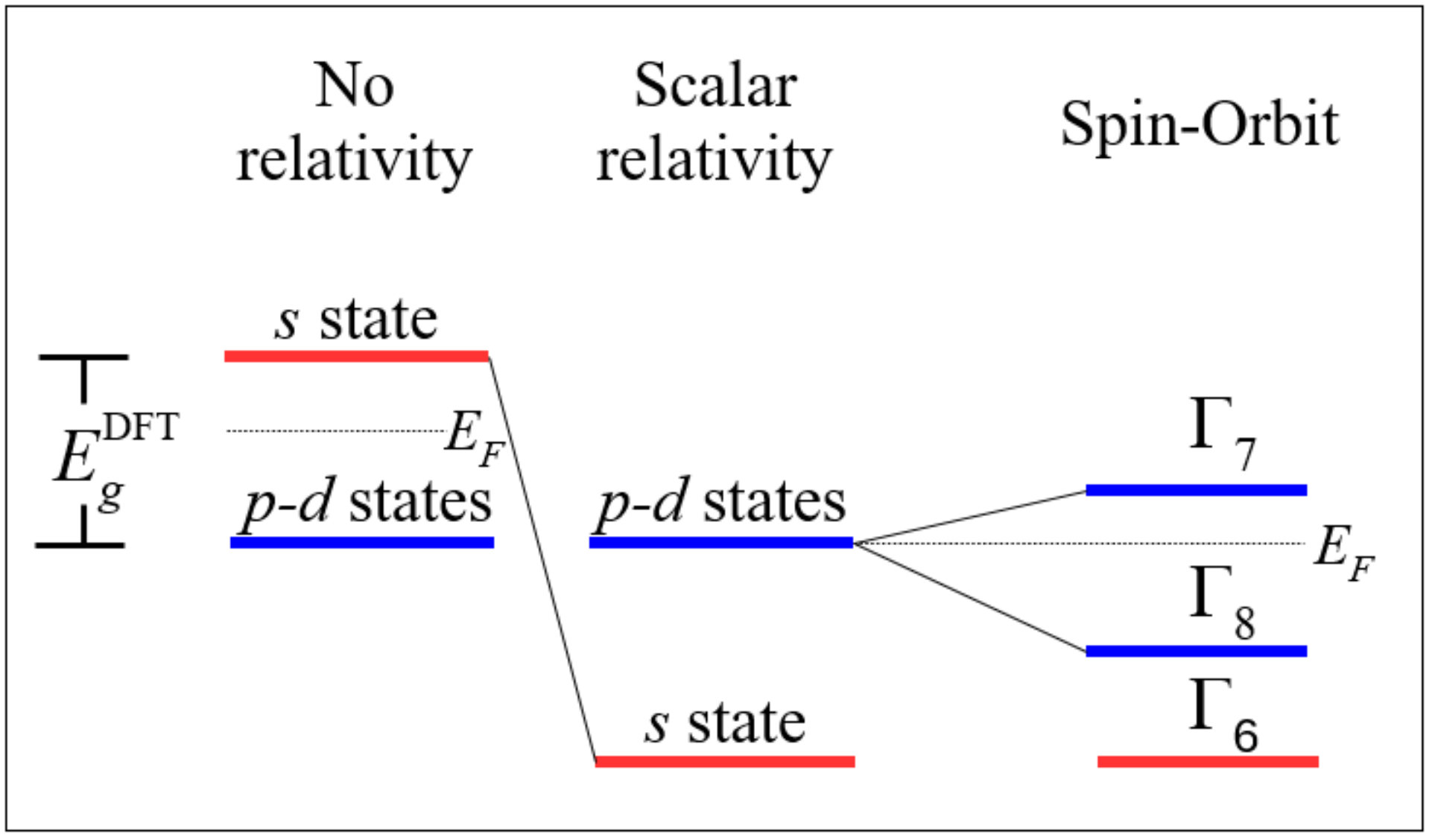}
    \label{fig:dft_schematic}   
    \end{subfigure}
    \begin{subfigure}[t]{0.49\textwidth}
    \centering
    \caption{}
    \includegraphics[width=\textwidth]{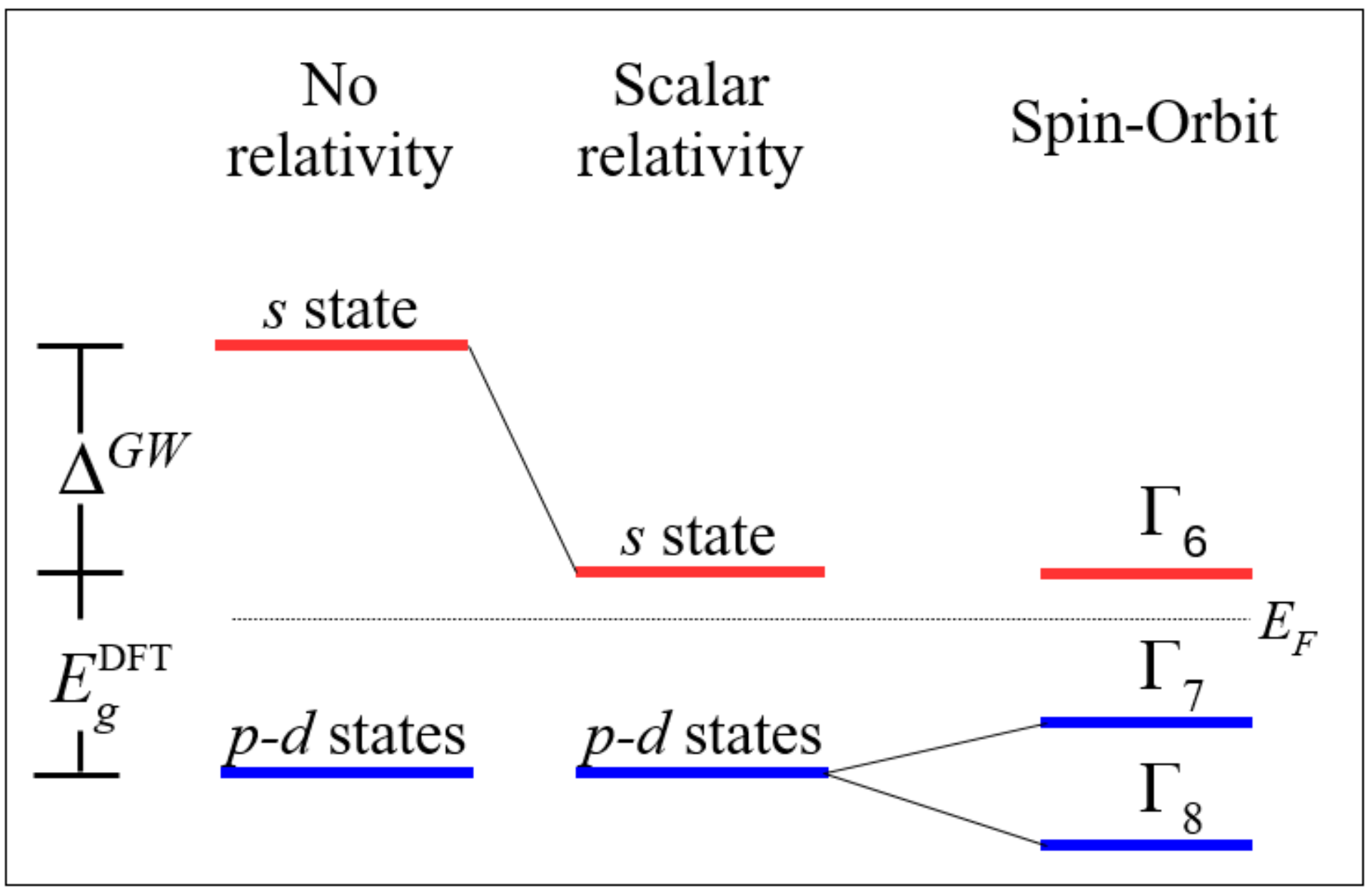}
    \label{fig:gw_schematic}
    \end{subfigure}
  \caption{(a) A schematic of the single-particle energies computed within DFT, first neglecting relativistic effects (left), then including scalar relativistic effects (middle), and spin-orbit coupling (right). The $p$-$d$ hybridized states are at the Fermi energy when including scalar relativistic effects.
  (b) A schematic of the single-particle energies when including corrections from $GW$ quasiparticle calculations. The $s$ state lowers in energy when including scalar relativistic effects but now remains above the Fermi energy.}
\end{figure*}

To better understand the ordering of the electronic states at the $\Gamma$ point, the bandstructure in DFT is calculated with non-relativistic and scalar-relativistic pseudopotentials. Scalar-relativistic pseudopotentials neglect spin-orbit coupling but include the physics of the Darwin term and the relativistic mass correction, which are both significant for Hg $s$ states. The non-relativistic pseudopotentials neglect these terms. Fig. \ref{fig:nonrel_dft} shows that the non-relativistic pseudopotentials give a bandstructure typical for a zincblende material with light atoms, where an $s$-like $\Gamma_1$ state is placed $E_{g}$ above occupied $\Gamma_{15}$ states. Incorporating the Darwin and relativistic mass-correction terms, however, the $\Gamma_1$ lowers below the Fermi energy (Fig. \ref{fig:scalar_dft}). The $p$-$d$ hybridized $\Gamma_{15}$ states then split to form parabolic conduction and valence bands that are degenerate at $\Gamma$, indicating a zero-gap semimetallic state. Incorporating spin-orbit coupling breaks this degeneracy, and the states can be identified as belonging to either the $\Gamma_7$ or $\Gamma_8$ representations. The schematic of the changing ordering of the energies upon inclusion of scalar relativistic effects and spin-orbit coupling is shown in Fig. \ref{fig:dft_schematic}.

Adding the quasiparticle energy correction to the band gap, $\Delta^{GW}$ (Fig. \ref{fig:gw_schematic}), places the $\Gamma_1$ state high above the Fermi energy such that scalar relativistic effects do not lower the state below the Fermi energy.
Turning on spin-orbit coupling then splits the occupied $\Gamma_{15}$ states into the $\Gamma_7$ and $\Gamma_8$ states, with the $\Gamma_7$ state higher in energy.
%(and we use the double-group notation $\Gamma_6$ for the $s$-like state)
,
% leading to the appropriate value of the band gap.
% We note that, in practice, by beginning with states and eigenvalues from FR-DFT, we have incorporated all relativistic effects first and then opened the gap with FR-$GW$.
%, with the physics of the final bandstructure being insensitive to the ordering of which process comes first when performed consistently.

%%========================================================================
\section{Conclusion}
\label{sec:conclusion_3}
%%========================================================================

The bandstructure of $\beta$-HgS near the Fermi energy
calculated by the the FR-$GW$ approach indicates that the order of the states differs from that predicted by fully-relativistic DFT calculations.
While DFT gives an ordering of the bands
$\Gamma_{6}$, $\Gamma_{8}$, $\Gamma_{7}$,
the present FR-$GW$ quasiparticle energies at the $\Gamma$-point is ordered
$\Gamma_{8}$, $\Gamma_{7}$, $\Gamma_{6}$.
This ordering of the states indicates a topologically trivial narrow band gap semiconductor.

The value of the band gap calculated within FR-$GW$ is 0.10~{eV}, which compares
favorably to the value of 0.11~{eV} from experiment\cite{shubnikov},
and the calculated value of the electron effective mass is 0.07~$m_e$,
which agrees well with the range of values cited in experiment of 0.04-0.07~$m_e$\cite{zallen,shubnikov}.
%This close agreement with experiment supports the conclusion that our presented calculation is sufficiently accurate.
The ordering is reproduced with other approaches of treating the frequency-dependence in the one-shot $GW$ approach, the COHSEX and GPP methods. The possible deviation of the Kohn-Sham orbital basis from the true quasiparticle states, estimated by the COHSEX approximation to the self-energy, indicates that the use of the Kohn-Sham states is accurate to within 0.1 percent.
%, so self-consistency in the quasiparticle wavefunctions would not appreciably change our results.

The reversal of the band inversion predicted in DFT can be understood in terms of the band gap problem. The underestimation of the band gap places the $s$-like $\Gamma_6$ state too low in energy, such that it becomes occupied due to the strength of its change in energy upon incorporation of the relativistic mass correction and Darwin terms captured in scalar- and fully-relativistic pseudopotentials.
The correction to the band gap from $GW$ calculations raise its energy
sufficiently to place it above the Fermi level.

%========================================================================
\section{Acknowledgments}
\label{sec:acknowledgments}
%%========================================================================
The authors would like to thank Felipe H. da Jornada, Mauro Del Ben, and Jack Deslippe for useful discussions.
This work was supported by National Science Foundation Grant No. DMR10-1006184 and by the Director, Office of Science, Office of Basic Energy Sciences, Materials Sciences and Engineering Division, U.S. Department of Energy under Contract No. DE- AC02-05CH11231. 
within the Nanomachines Program (KC1203) and within the  Theory of Materials Program (KC2301).
Computational resources have been provided by the DOE at Lawrence Berkeley National Laboratory's NERSC facility.

\bibliography{full_bib}

\begin{figure}[p]
  \centering
  \includegraphics[width=0.5\textwidth]{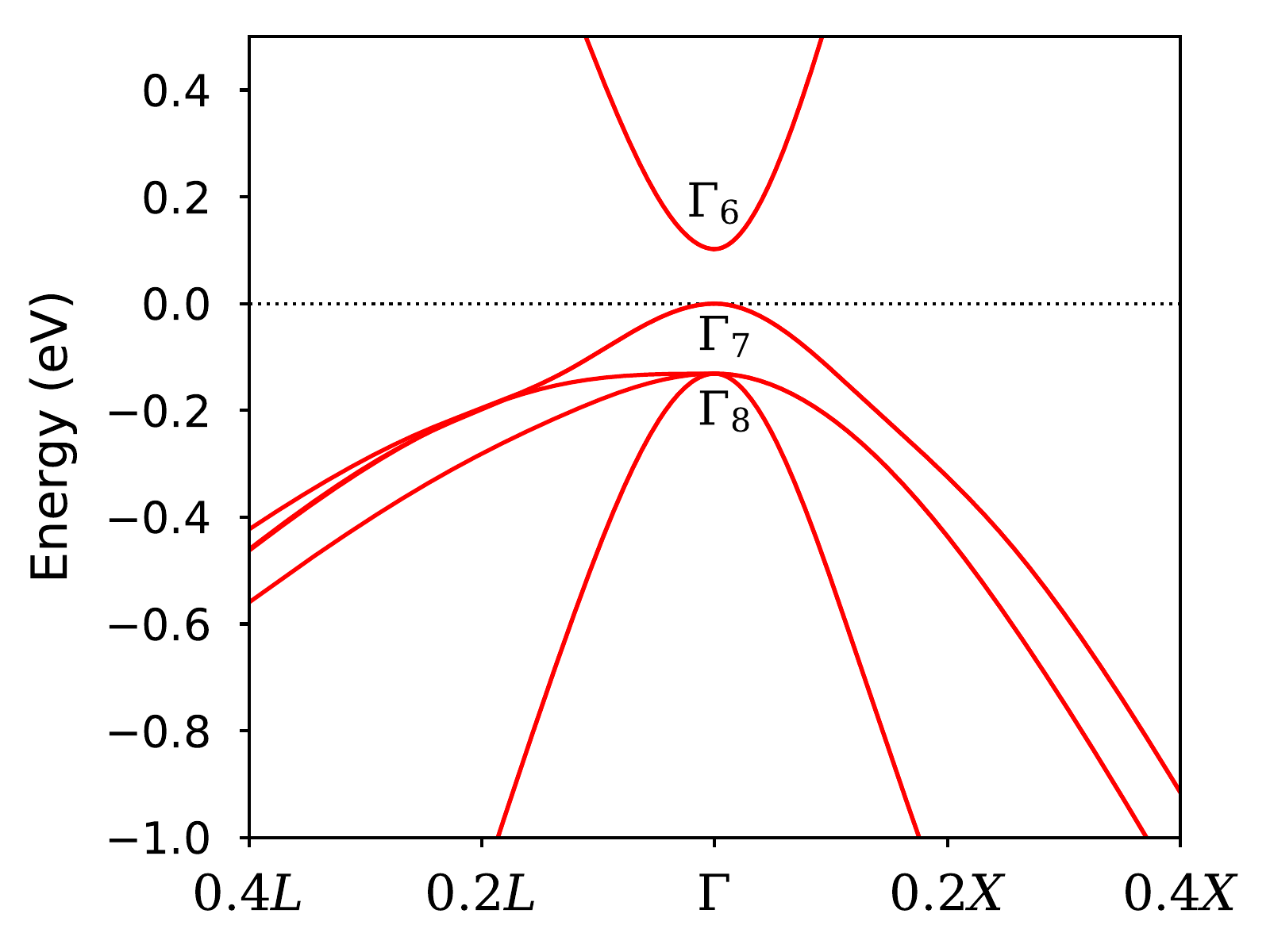}
  \caption{The quasiparticle bandstructure of $\beta$-HgS, computed at the FR-$GW$ level using the contour deformation method. The states at the $\Gamma$-point, in increasing energy, are $\Gamma_8$, $\Gamma_7$, $\Gamma_6$.}
  \label{fig:eqp}
\end{figure}

\begin{table}[p]
\caption{
A comparison of the results for interband gaps near the Fermi energy from the literature.
}
\begin{tabular*}{\textwidth}
{c @{\extracolsep{\fill}} lccccc}
\hline\hline
  & $E_0$ (eV) & $\Delta^{\textrm{SOC}}$ (eV) & $E_g$ (eV) & Basis set \\
\hline
Present Work (FR-$GW$)                        &  0.23  &  -0.13  &  0.10   &  Plane-wave (PW)\\
$GW$+SOC\footnote{\label{a}Ref. \onlinecite{fleszar}} &  0.18  &  -0.12  &  0.06   &  Gaussian+PW\\
hybrid-QSGW\footnote{\label{b}Ref. \onlinecite{van_schilfgaarde}}  &  0.37  &  -0.07  &  0.31   &  LMTO \\
FR-$GW$\footnote{\label{c}Ref. \onlinecite{sakuma}}  & -0.02  &  -0.19  &  0.02   &  FLAPW \\
FR-$GW$\footnote{\label{d}Ref. \onlinecite{west}}  & -0.02  &  -0.10  &  0.02   &  PW \\
\hline\hline
\end{tabular*}

%Comparison of results for interband gaps near the Fermi energy from the literature.
%All calculations used the LDA for the DFT calculations and captured the frequency dependence of the screening
%and self-energy with the exp
%licit calculation of the frequency dependence, either with real frequencies or the 
%use of the contour deformation method with imaginary frequencies.
%$^a$ Ref. \cite{fleszar},
%$^b$ Ref. \cite{van_schilfgaarde},
%$^c$ Ref. \cite{aryasetiawan},
%$^d$ Ref. \cite{west}.
%}
\label{tab:table2}
\end{table}

\begin{table}[p]
\caption{\label{tab:table1}
Interband gaps near the Fermi energy for $\beta$-HgS. From DFT, the four-fold
$\Gamma_8$ states are higher in energy than the two-fold $s$-like $\Gamma_6$ states. With the inclusion
of self-energy effects (COHSEX, GPP, or Contour deformation), the $\Gamma_8$ states are lower in energy
than the now-unoccupied $\Gamma_6$ states.}
\centering
\begin{tabular*}{0.65\textwidth}
{c @{\extracolsep{\fill}} lcccc}
\hline
  & $E_0$ (eV) & $\Delta^{\textrm{SOC}}$ (eV) & $E_g$ (eV) \\
\hline\hline
DFT             & -0.54  &  -0.10  &  0.10 \\
COHSEX          &  0.92  &  -0.12  &  0.80 \\
GPP             &  0.37  &  -0.15  &  0.22 \\
Contour deformation  &  0.23  &  -0.13  &  0.10 \\
\hline\hline
\end{tabular*}

\end{table}

\end{document}